# An Auditability, Transparent, and Privacy-Preserving for Supply Chain Traceability Based on Blockchain

**Bora Buğra SEZER[1], Selçuk TOPAL[2], Urfat NURİYEV[1]**

[1]*Department of Mathematics, Ege University, İzmir, TURKEY*
[2]*Department of Mathematics, Bitlis Eren University, Bitlis, TURKEY*

[bora.bugra.sezer@ege.edu.tr](mailto:bora.bugra.sezer@ege.edu.tr), [s.topal@beu.edu.tr](mailto:s.topal@beu.edu.tr), [urfat.nuriyev@ege.edu.tr](mailto:urfat.nuriyev@ege.edu.tr)

Abstract

Traceability and auditability are key structures that are vital in supply chain management and construction. However, trust is the most important aspect of customers in these systems. Also, we have to rely on third parties to trade in centralized systems. Although current exist frameworks for these solutions in the supply chain, these have work poor traceability and lack of real-time information, and especially lack of privacy-preserving. In this paper, we propose a privacy-preserving framework for supply chain traceability that using smart contracts. Development processes, model implementation, and smart contracts are presented in detail and we show cryptographic techniques in these technologies to address the aforementioned. Finally, thanks to traceability and auditability, customers and other parties can view with a single product ID and also verified with digital signature the claims of the actors in the system.

Highlights

- Privacy-preserving has provided for every actor from producer to consumer.
- With a single product ID, the consumer can view the entire stage.
- Digital signing has been performed in off-chain.
- Authentication has been performed with the elliptic curve algorithm in the contract.
- For scalability, both event-based contracts have been written and necessary operations have been performed off-chain.

## 1. Introduction

Supply chain (SC) covers the process from the supplier of all products and services to the customer at the last stage. It can be explained as the name of the concept that covers all activities, human resources, technology, company structures and resources on this path. All kinds of resources and components involved in all activities carried out in the supply chain process are converted into products and delivered to the customer at the last stage. Within the body of relationships and connections that make up SC, there is a movement towards suppliers, distributors, wholesalers, retailers and finally consumers. Each link of this chain regularly follows each other. Similarly, we can say that SC has certain stages in terms of business processes. Valuable processes such as production, stock management, material supply, distribution, sales, procurement, customer relations management ensure the functioning of SC. Blockchain offers another alternative that can bring together different parties that have not directly established trusting relationships with each other through the transparency it provides while there are other technology options available to help manage Ss. Blockchain stores every transaction or data exchange that takes place on the network, potentially reducing the need for third parties and/or agents, by providing a way that all parties in the network can share access to the same data, including those added to data chronologically. Data in Blockchain cannot be removed or mutable. By allowing each party to see the same data in near real time, the blockchain can help eliminate the complex and costly data consensus required by most systems in today's world. It allows both privacy and transparency to declare and broadcast data both globally and locally while sharing data with relevant parties without any doubt.

Traceability and auditability have begun to very important structures in SC which has become more complex and expanded in the last decades. SC traceability is the process of tracking the origin and voyage of products and their inputs from the very beginning of SC to expiration. Traceability considers individual component groups or purchase orders as they move through the supply chain. The specificity of the data used in traceability gives permission more intended recalls, reducing scale and cost to businesses. SC auditability is an emerging study area in the SC and Blockchain. Data storage, transactions and consistency rules in a Blockchain are vital points for auditability. Integrating Auditability's functions within the Blockchain with SC and transforming it into a transparent service requires high attention, detailed security studies and scalability.

Also, traceability and privacy are important for the performance and security of the enterprise, and more importantly for regulations (international), and their importance is increasing day by day.    For traceability and auditability in SC, a system that records and tracing of products is needed. Various traceability standards such as RFID (Radio-Frequency Identification) [9], EPC (Electronic Product Code) global traceability standards have also been established for this tracing process [2]. Also, the current SC requires trust in third parties. The fact that there is trust in a central system and that third parties control it makes this situation difficult due to the nature of SC. Therefore, distributed systems started to be built and the traceability, security, auditability of the existing system has been limited.

Blockchain technology is a crucial technology that emerged with Bitcoin in 2009 [1], eliminates trust in third parties, provides trust in a trustless environment, and implements it without a third party. After the emergence of Bitcoin, this technology has been applied to various industries and successful results have been obtained in finance [3]. Various studies have been carried out to ensure traceability in SC [4-7]. Recently, interest in blockchain technology has increased and researchers have applied the supply chain system to this technology. However, most of these studies have been proposed either theoretically or for a centralized system [5-6] and also does not solve the problems related to privacy-preserving.

In this paper, we propose a framework that ensures the confidentiality and traceability of the actors and transactions involved in the system and auditability for consumers. Digital signing has been performed with elliptic curve cryptography [8] that used public key infrastructure for the confidentiality of actors and transactions. Where necessary, the relevant actor can request the private key so that the product can be tracked securely.

The rest of the paper is organized as follows: we briefly introduce blockchain technology and provide a preliminary assessment of our work by reviewing previous studies in the relevant field. Afterward, we introduce the framework. We then analyze in detail the traceability and auditability of our system with the experimental results. We discuss the privacy-preserving of our system that is how public-key cryptography is used for verification. Finally, our discussion and conclusion are presented.

## 2. Literature Review

After the successful use of blockchain technology in cryptocurrency systems, various research has been carried out for the use of this technology in different areas. Researches have been primarily on transaction privacy. In other words, unlike the existing decentralized systems, it has shown that financial transactions cannot be stored clear on the blockchain, and transaction privacy is public's protected [10].

Another study obtained using cryptographic techniques is ZCash. k-SNARKs (used by ZCash) is a non-interactive zero-knowledge proof of knowledge. However, the desired efficiency could not be obtained due to reasons such as high cost and limitation of performance [19].

In recent years, traceability has started to attract attention in sectors such as food, pharmacy, and non-durable agricultural food products [13-16]. Also, important academic work has been done in terms of traceability to trace the products [17-18]. Other studies have also made applications of RFID technology in the supply chain which the use of advanced technology. In their studies, it has been shown that the use of RFID technology in the supply chain provides coexistence between actors and is more efficient [11].

However, it has been claimed that RFID technology can be solved with a central system in the supply chain. Although the tags used in this technology are combined with blockchain technology, separate aspects of blockchain technology and separate aspects of traceability has been discussed. Performance, integrity, and applicability claims do not confirm during the analysis process in which the studies showed [12].

When studies are investigated that using blockchain technology, it has been observed data privacy sensitivity do not consider. Besides, auditability for actors and customers is not covered in detail. As a result of the researches on the parameters of transparency, integrity, auditability, and traceability, this is not done by considering the integrity of SC.

## 3. Conceptual Framework

3.1. Preliminaries

Blockchain technology is a distributed technology that does not need a central authority. This technology, which is a peer-to-peer system, originally emerged with Bitcoin [1] to support cryptocurrency. And the security of the system depends on strong cryptographic schemes that connect and validate transactions in blocks to the Merkle root [1] (Figure 1).

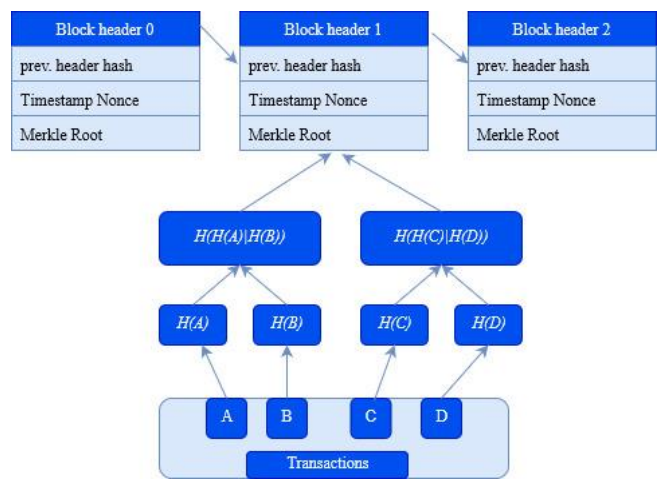

Fig. 1. Blockchain Transactions (Simplified).

Smart contracts used in blockchain technology first introduced by Nick Szabo [23]. Thanks to smart contracts, the relevant rules have entered into the system without a third party. A part of our framework has built on-chain using solidity programming in Ethereum Virtual Machine (EVM) that supports smart contracts [24] (Fig. 2).

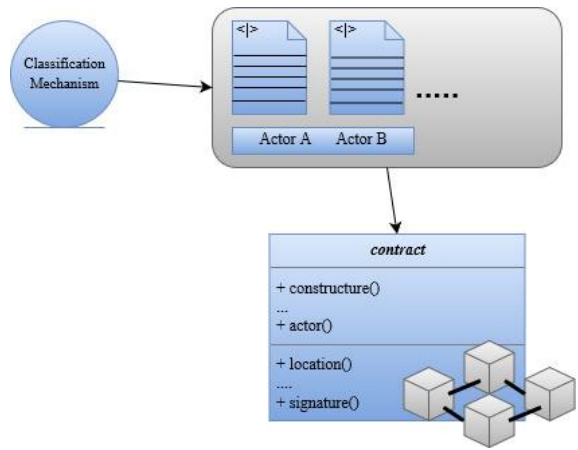

Fig. 2. Smart Contracts diagram.

Public key cryptography uses a public-private key pair to encrypt and decrypt data. Since larger keys are used in the RSA [30] public-key encryption system, a NIST [21] approved elliptic curve digital signature algorithm (ECDSA) has been proposed to overcome this disadvantage [20]. The digital signatures have integrity non-repudiation and authentication. Therefore, in our study, the digital signing has performed using the elliptic curve encryption algorithm in smart contracts. [22].

3.2. Model

We propose a system model using smart contracts based on a public-permission blockchain that is private assets in the public network. Our model consists of 5 actors: Supplier, Producer, Dealer, Retailer, Customer.

Each type of product has managed in our system is processed by the actors and these are realized as on-chain within a smart contract. The digital signing has been performed which the most important structure of our frame in an off-chain. Therefore, using off-chain has been better performed for the scalability. After the digital signing process, the verification has done in the contract. we show that in figure 3, the supplier creates the product and sends it to the producer via transport. The producer is then forwarded to the retailer and the retailer is forwarded to the dealer and final customer.

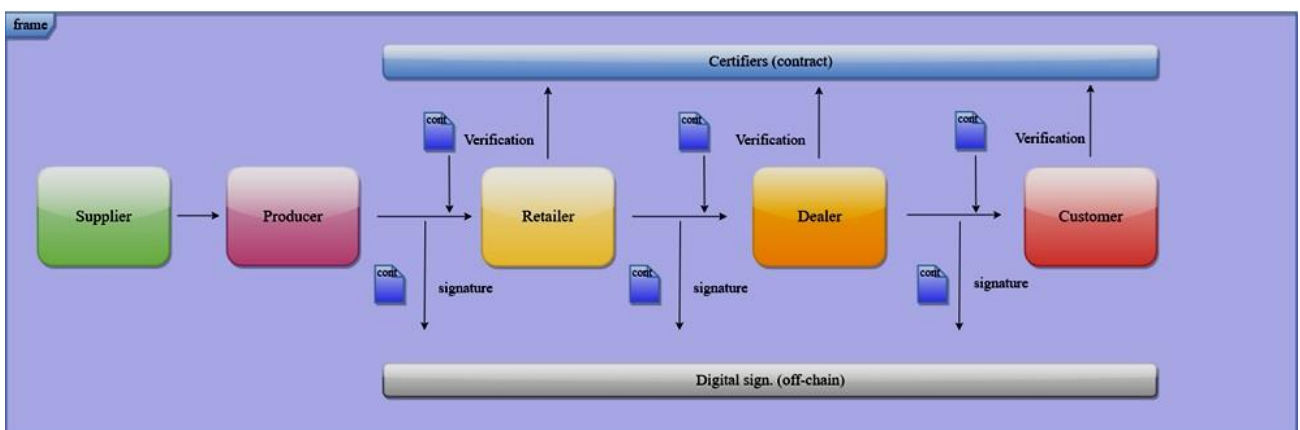

Fig. 3. The conceptual framework of the supply chain system that includes privacy-preserving.

Product information is created by the actors in the system and transactions are broadcast to the nodes in the network. Then the transactions are validated by all nodes in the network.

3.3. Details of transactions of the frame

Each actor in the system has its own actor id's, where denoted $a_{id}$. Then, there is a product id denoted $p_{id}$ for each product. The product information for each product denoted $p_{inf}$, the hash value of the relevant product's data denoted $h(ptx_i)$, the date and time of the product purchase-sale transaction denoted $p_{intime}$, $p_{outtime}$, , the hash value of the previous transaction denoted $h(ptx_{prev.})$, the digital signature of the relevant actor and the verification process denoted $sign(h(ptx_i))$, $verfy(sign(h(ptx_i)), sig_h(h(ptx_i)))$ (Table 1).

**Table 1**
Structure of the frame

| Symbol | Description |
| --- | --- |
| $a_{id}$ | Actor id |
| $p_{id}$ | Product id |
| $p_{inf}$ | Product information (pcount etc.) |
| $h(ptx_i)$ | Hash of Product |
| $p_{intime}$, $p_{outtime}$ | Time of transaction |
| $h(ptx_{prev.})$ | Hash of prev. transaction |
| $sig_h(h(ptx_i))$ | Sign hash of hash |
| $sign(h(ptx_i))$ | Signature of actor |
| $verfy(sig(h(ptx_i)), sig_h(h(ptx_i)))$ | Verification of actor |

Where the transaction hash $h(ptx_i) = \{a_{id}, p_{id}, p_{inf}, p_{intime}, p_{outtime}, h(ptx_{prev.})\}$.

3.3.1. Signature

According to the public key cryptography elliptic curve algorithm which is used in the digital signature, each actor has a public-private key pair. the digital signature has been performed outside of the contract (off-chain). Signing is shown in alg1.

> **Pseudocode : Algorithm 1** *Signature of transactions*
> *function sign_actor*
> *check $a_{id} \neq 0$, if not ret. false*
> *check $p_{id} \neq 0$, if not ret. false*
> *$h(ptx_i) \leftarrow hash(a_{id}, p_{id}, p_{inf}, p_{intime}, p_{outtime}, h(ptx_{prev.}))$*
> *$p_{intime} <$ current time, if not ret. false*
> *$sign(h(ptx_i))$ is valid, if not ret. false*
> *return true.*
> *end function*

3.3.2. Authentication

The digital signature has been used in our frame to prove the validity of transactions, to protect confidentiality, and to prevent fraud. Signing has been performed that the hash value of each transaction by the actor. Then the signed hash value of the product's hash value is generated in the contract. Authentication has been performed used the signature of the actor and the signed hash value of the products with the elliptic curve algorithm in the contract. Shown in algorithm 2.

**Pseudocode: Algorithm 2** *Validation of digital signature*

*function verfy_actor*
*check Sign_actor =true, if not ret. false*
*$sig_h(h(ptx_i))$ ←signature hash($h(ptx_i)$)*
*$s_{adr}$←sender actor adress*
*verfy($a_{adr}$) ←verify with ECDSA $sig(h(ptx_i))$, $sig_h(h(ptx_i))$*
*if vefy($a_{adr}$)≠ $s_{adr}$ then ret. false*
*end if*
*return true*
**end function**

## 4. Security Analysis

We summary a privacy-preserving in our frame from two aspects. The first is protecting the identity of the actors (anonymity) and the other is the focus on protecting the actors' transactions. The security of our framework is based on the elliptic curve cryptography. Elliptic curve cryptography is an encryption system based on the public key encryption system and also the algebraic structures of elliptic curves over a finite field.

The security of the transactions is ensured by using the keccak-256 ($h(ptx_i)$) [26]. The digital signature (ECDSA) $sign(h(ptx_i))$ has been performed by the actors in the off-chain.

**Definition 1**: Elliptic curve is a plane curve defined by the equation below (2.1).

$$y^2 = x^3 + ax + b, \text{ where } (x, y) \in Z_p \quad (2.1)$$

We need to define the arithmetic operations on an elliptic curve and the generation of a key pair. First of all, the basic operations between two points on the curve (p, q) are summed and the third point where r is formed is again on the curve (r = p + q). This process is illustrated by the formula below (2.2).

$$\lambda = (y_q - y_p)/(x_q - x_p)$$
$$x_r = \lambda^2 - x_p - x_q \quad (2.2)$$
$$y_r = \lambda(x_p - x_r) - y_p$$

The special case on the curve shown by the formula below is the addition of double point (2.3).

$$\lambda = (3x^2_p + a)/(2y_p) \quad (2.3)$$

Here, any *q* point on the curve is a public key, another point is a *g* generator point. Multiplier *k*, which is the factor of g is the private key (Table 2).

## Table 2
public-private key pair on the elliptic curve

| private key | k, $k \in R\ [1, n-1]$ |
|---|---|
| public key | $q \leftarrow kg$, g generate point |
| public-private key pair | (q,k) |

**Definition 2**: Let E elliptic curve over finite field $F_p$ such that $p,q \in E(F_p)$ two points and $k \in R\ [1, n – 1]$ such that $p = kq$. It is impossible to find k in polynomial time (ECDLP).

Despite the simplicity of the arithmetic operations, solving and computing are difficult. It is not enough for an attacker to know the public key at the relevant point to solve the system. The attacker needs to know the place at the starting point and the multiplier. To find these values, the attacker has to test all the values. Even if the attacker knows the starting point, he has to extract the values from the public key point to the starting point he finds to find the multiplier that can obtain these values. This is not happening in polynomial time. Therefore, the security based on the Elliptic Curve Discrete Logarithm Problem (ECDLP).

**Lemma** In our frame, it is infeasible to determine the identity (ID) of any actor in polynomial time by an attacker.

**Proof Suppose** the attacker knows the $sign(h(ptx_i))$ signed by any actor. In order for the attacker to know the actor ID, he must also have the signed hash value $sig_h(h(ptx_i))$ and test these values. Due to the ECDLP and one-way hash function, it is not possible to find the identity ($a_{adr}$) of the actor. This situation provides anonymity.

As a result, we say that the security just can be discussed over the elliptic curve discrete logarithm problem (ECDLP) [8-25].

## 5. Complexity

When calculating the complexity in our framework, we need the number of transactions performed by the actors. In our framework, the main point has the transactions created by the actors and the digital signature of these transactions. Also another point, the verification of transactions performed after a digital signature by actors.

The complexity of the elliptic curve algorithm used for the signing and the verification is linear that depends on key-length. Verification, since it depends on the digital signature, its complexity depends on the key-length and is linear. Let the key-size of the elliptic curve be k (Table 3).

## Table 3
Complexity of the framework

| symbol | description | Details (protocols of the frame) | Complexity |
|---|---|---|---|
| $A_k$ | number of transactions (for each actor) | Transactons created(contract) | $O(A_k)$ |
| $S_k$ | Signature by each actor | Digital sigtanure (off-chain) | $O(S_k)$ |
| $V_k$ | Validation by each actor | Validation (contract) | $O(V_k)$ |

we assume that there are N actors in the network. The transactions, signing, and verification created by each actor are broadcast to all other actors in the system. Then, we can say that the number of transactions increases linearly as depends on key-size. Therefore, the fact that the complexity is linear means that our framework performs in terms of scalability. As seen in Table 2, the complexity is *O(N) that* the number of transactions increases linearly.

## 6. Experimental Results

In our proposed framework, information of the product is traceable and auditable from producer to consumer. Any actor in the system can be performed verification and traceability with product ID. It is assumed that the actors have an address of their own.

To achieve the applicability of our frame, we used truffle-ganache-cli, web3 javascript programming, and web-based (HTML). Our work executed on a local computer running ubuntu 18.04. We created a proof of concept that facilitates the examination. In order to realize the proposed framework, we have blended the components defined with the existing smart contract. We wrote our smart contract that are deployable on compatible blockchain in Ethereum [27](EVM).
EVM has a gas fee for each transaction. In EVM, the gas cost is an expensive process. Also using the details and storage in on-chain increases the gas fee. Increasing gas costs cause negativity in scalability.

Scalability depends on many variables such as public or permissioned ledger usage, accessibility, and privacy. In the use of a public ledger, scalability depends heavily on gas costs. Firstly, we need to use ethereum's web3 javascript API[1] to add a new product and run it on off-chain. Therefore, we used events in off-chain web3-API[1] rather than storing them in the contract. Also, we used web-based tools to easily manage the studies. Even if there is no storage in the use of events, they can be easily observed and verified on the blockchain.

In our contract, we separate the costs of deploying and execution costs. Since certain product features are required to be executed in the contract, we stored them in the contract. Also, we compared the gas fee both by using the events and by performing all the transactions on the contract (Figure 4). As can be seen in figure 4, gas fees increase linearly in both approaches, but the use of events-based web3-API[1] has significantly decreased the gas fee.

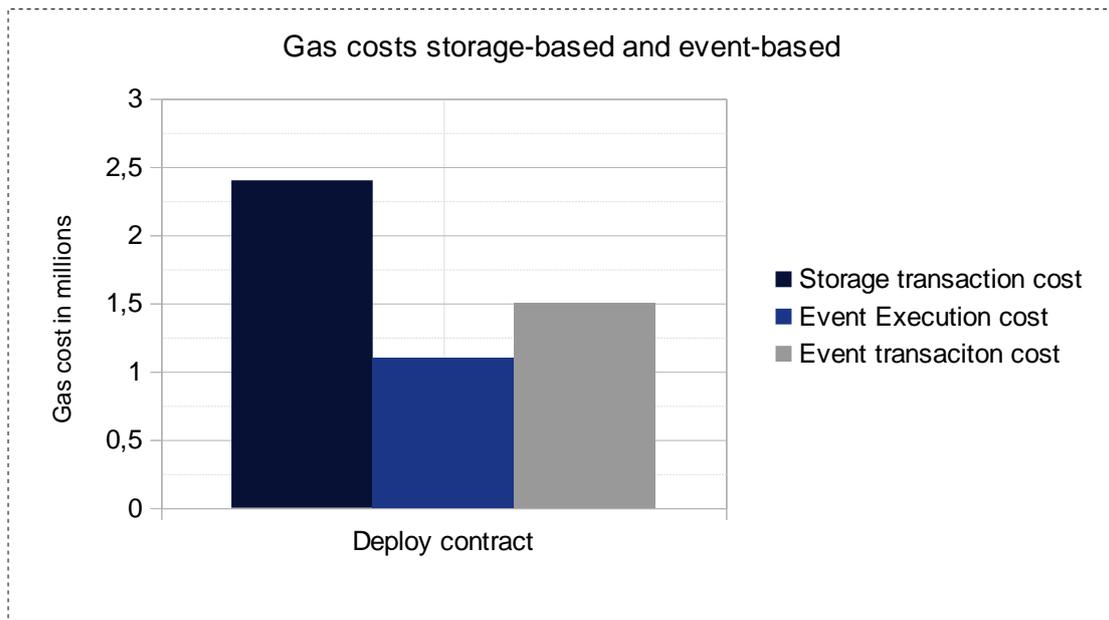

Fig. 4 Average gas costs of the contract by inputs and application.

For digital signing and other cryptographic transactions, 256-bit key length used which considered safe by NIST[2]. It reduced the workload on the contract by using off-chain. Since the transactions on the blockchain are immutable, fraud and alteration do not be possible.

Recently, Ethereum has switched to proof-of-stake consensus protocol with 2.0. Also, there are suggestions that the new consensus protocol developed such as Algorand [28], Avalanche [29] results in up to 100 times transactions per second (tps). Therefore, our framework can be used safely with these consensus protocols. As a result of the experiments, our proposed frame has provided real-time traceability.

## 7. Conclusion

In this paper, we have prepared a framework for the supply chain that is traceable, auditable, privacy-preserving using smart contracts. Each actor signs hash of the transaction using a digital signature in the off-chain.
Thus, the integrity and non-repudiation of the system are ensured. The consumer can see and control the products transparently with a single productID.

The actor verifies using elliptic curve public key encryption. therefore, we ensure data confidentiality between actors. This provides privacy-preserving in our framework. As a result, an applicable, auditable supply chain model that includes privacy-preserving on blockchain technology has been successfully achieved with experimental results. This will increase the safety of the products in the supply chain, as it provides trust in a trustless environment.